\documentclass{aa}  
\usepackage{graphicx}
\usepackage{txfonts}
\def\Msun{M_\odot}

\def\cmtres{\relax \ifmmode {\,\mbox{cm}}^{-3}\else \,\mbox{cm}$^{-3}$\fi}
\def\cmdos{\relax \ifmmode {\,\mbox{cm}}^{-2}\else \,\mbox{cm}$^{-2}$\fi}
\def\cmseis{\relax \ifmmode {\,\mbox{cm}}^{-6}\else \,\mbox{cm}$^{-6}$\fi}
\def\ergs{\relax \ifmmode {\,\mbox{erg\,s}}^{-1}\else \,\mbox{erg\,s}$^{-1}$\fi}
\def\kms{\relax \ifmmode {\,\mbox{km\,s}}^{-1}\else \,\mbox{km\,s}$^{-1}$\fi}
\def\ha{\relax \ifmmode {\mbox H}\alpha\else H$\alpha$\fi}
\def\hb{\relax \ifmmode {\mbox H}\beta\else H$\beta$\fi}
\def\hi{\relax \ifmmode {\mbox H\,{\scshape i}}\else H\,{\scshape i}\fi}
\def\hii{\relax \ifmmode {\mbox H\,{\scshape ii}}\else H\,{\scshape ii}\fi}
\def\oiii{\relax \ifmmode {\mbox O\,{\scshape iii}}\else O\,{\scshape iii}\fi}
\def\oii{\relax \ifmmode {\mbox O\,{\scshape ii}}\else O\,{\scshape ii}\fi}
\def\oi{\relax \ifmmode {\mbox O\,{\scshape i}}\else O\,{\scshape i}\fi}
\def\nii{\relax \ifmmode {\mbox N\,{\scshape ii}}\else N\,{\scshape ii}\fi}
\def\sii{\relax \ifmmode {\mbox S\,{\scshape ii}}\else S\,{\scshape ii}\fi}
\def\lha{\relax \ifmmode \mbox {L}_{H\alpha}\else $\mbox{L}_{H\alpha}$\fi}
\def\ldig{\relax \ifmmode {\mbox L}_{DIG}\else ${\mbox L}_{DIG}$\fi}
\def\ls{\relax \ifmmode {\mbox L}_{ Str}\else ${\mbox L}_{ Str}$\fi}
\def\eme{\relax \ifmmode {\,\mbox{pc\,cm}}^{-6}\else \,pc\,cm$^{-6}$\fi}
\def\l{\relax \ifmmode  \lambda\else $\lambda$\fi}

\def\etal{{et al.~}}
\def\me{$^{-1}$}
\def\arcmin{\hbox{$^\prime$}}
\def\arcsec{\hbox{$^{\prime\prime}$}}
\def\deg{\hbox{$^\circ$}}
\def\fs{\hbox{$^{\rm s}$}}
\def\hms#1h#2m#3s{\relax \ifmmode #1^{\rm h}\,#2^{\rm m}\,#3^{\rm s}
                   \else \hbox{$#1^{\rm h}\,#2^{\rm m}\,#3^{\rm s}$}
                  \fi}
\def\dms#1d#2m#3s{\relax#1\deg\,#2\arcmin\,#3\arcsec}
\def\hmsd#1h#2m#3.#4s{\relax\ifmmode #1^{\rm h}\,#2^{\rm m}\,#3.#4\fs
                      \else \hbox{$#1^{\rm h}\,#2^{\rm m}\,#3#4\fs$}
                      \fi}
\begin{document}

   \title{An Evolutionary Sequence of Expanding Hydrogen Shells in Galaxy Discs}

   \author{M. Rela\~no
          \inst{1}
          \and
          J. E. Beckman\inst{2,3}
         \and
          O. Daigle\inst{4}
        \and
          C. Carignan\inst{4}
          }

   \offprints{M. Rela\~no}

   \institute{Dpto. F\'\i sica Te\'orica y del Cosmos,
              Universidad de Granada, Avda. Fuentenueva s/n, 18071, Granada, Spain \\
             \email{mrelano@ugr.es}
         \and
             Instituto de Astrof\'\i sica de Canarias, C. V\'\i a
              L\'actea s/n, 38200, La Laguna, Tenerife, Spain
         \and
             Consejo Superior de Investigaciones Cient\'\i ficas
              (CSIC), Spain \\
                          \email{jeb@ll.iac.es}
         \and
             Observatoire du mont M\'egantic, LAE, Universit\'e de
              Montr\'eal, C. P. 6128 succ. centre ville, Montr\'eal, Qu\'ebec, Canada H3C 3J7 \\
                            \email{odaigle@ASTRO.UMontreal.CA,carignan@ASTRO.UMontreal.CA}
             }

   \date{Received ; accepted}

% \abstract{}{}{}{}{} 
% 5 {} token are mandatory

  \abstract
   {}
   {Large HI shells, with diameters of hundreds of pc and expansion 
velocities of 10-20~\kms\ have been detected in their
  hundreds in the Milky Way and are well observed features of local gas rich
    galaxies. These shells could well be predicted as a result of the
    impact of OB associations on the ISM, but doubt has been cast on this
    scenario by the apparent absence of OB stars close to the centres of a
    large fraction of these shells in recent observations of the
  SMC. Here we present observational evidence within a energetically 
consistent framework which strongly supports the scenario in which OB 
associations do produce the giant HI shells.}
   {Using Fabry-Perot scanned \ha\
    emission line mapping of nearby galaxy discs, we have detected, in
    all the \hii\ regions where the observations yield sufficient
    angular resolution and S:N ratio, dominant \ha\ shells with radii a few tens of pc, expanding
    at velocities of 50-100~\kms, and with gas masses of $\rm 10^4-10^5\Msun$. In previous studies, we found
    that stellar winds alone can account for the energetics of most of  
    the \ha\ shells, which form initially before the stars explode as
    SNe. We have applied a simple dynamically 
consistent framework in which we can extrapolate the properties of the
observed \ha\ shells to a few $10^7$yr after the formation of the OB stars.
The framework includes the dynamical inputs of both winds and SNe on the 
surrounding ISM. The results give quantitative statistical support to the 
hypothesis that the \ha\ emitting shells are generic progenitors of the 
HI shells.}
   {The results are in good agreement with the ranges of masses
  ($\sim$$\rm 10^6\Msun$), velocities (up to $\sim$20~\kms), and
  diameters (up to $\sim$500 pc) of representative HI shells observed in
    nearby galaxies. The combined effects of stellar winds, acting during
    the first few $10^6$yr, and SN explosions, "switching on"
    subsequently, are required to yield the observed parameters.}
   {}


\titlerunning{Evolutionary Sequence of Hydrogen Shells}
\authorrunning{Rela\~no et al.}

   \maketitle

%
%________________________________________________________________

\section{Introduction}

The presence of neutral hydrogen shells in the gas of disc galaxies 
is a widely known and observed phenomenon. Starting over 25 years 
ago Heiles (1979, 1984, 1990) detected large shell structures in 
Galactic neutral hydrogen, and in the intervening period a number of 
authors (e.g. Brinks 1981, Brinks \& Bajaja 1986, Deul \& den Hartog 1990, 
Puche et al. 1992, Kamphuis \& Sancisi 1993, Stavely-Smith et al. 1997, 
Wilcox \& Miller 1998, Walter \& Brinks 1999, Kim et al. 1999, 
Wills et al. 2002) have shown that such shells, often detected around 
holes in HI, are present in nearby gas rich galaxies. More recently 
Mclure-Griffiths et al. (2002) mapped hundreds of HI shells within the 
Galaxy, and Ehlerova \& Palous (2005) showed that in the outer galaxy 
these shells occupy a minimum of 5\% of the disc volume. Although telescope 
resolution limits the minimum detectable size of an HI shell to some tens 
of pc even in nearby galaxies, many have been detected on scales of 
hundreds of pc, up to 1~kpc in diameter.

Brinks \& Bajaja showed that in M31 HI holes smaller than 
300 pc are correlated with OB associations, but no reliable correlations 
could be found for bigger holes. However, when sufficient spatial 
resolution can be applied, such as in the case of the SMC, it is 
notable that positional coincidence of the HI shell with a corresponding 
OB association occurs in only a minority of cases, as shown by 
Hatzidimitriou et al. (2005). These authors showed that of the 509 
expanding HI shells catalogued in the SMC only some 40\% are spatially 
correlated with OB associations, and some 10\% lie in low density fields 
with no associated young stellar objects. Rhode et al. (1999) 
found evidence of young stellar clusters in only a small 
fraction of the HI holes in the dwarf galaxy Ho II. The age of the central 
stellar cluster located within the supergiant shell in the dwarf irregular
galaxy IC 2574 is in agreement with the kinematic age of the shell, supporting the 
scenario in which the supernovae and stellar winds create the giant HI holes 
(Stewart \& Walter 2000).

Mechanisms of different kinds for producing large HI shells 
have been presented in the literature. The possibility that 
the infall of a high velocity cloud can give rise to HI holes 
and shells has been discussed by a number of authors with some 
quite detailed models (e.g. Tenorio-Tagle \& 
Bodenheimer 1988, Rand \& Stone 1996, Santillan et al. 1999, 
Murray \& Lin 2004). In the last two of these studies magnetic 
fields were also included. Another possibility, proposed by 
Hatzidimitriou et al. (2005) is a turbulent origin for at least 
some of these structures. The present article does not claim the 
scope to contrast those theoretical models with observations. All 
we will do is to present our scenario, based on observations of 
expanding \ha\ shells in \hii\ regions, which offer a new quantitative 
element capable of contributing to explain the range of parameters 
observed for HI shells, including the largest of them.

The essential point of this article is to present evidence, using a set of 
observational data not hitherto applicable to this scenario (that of the 
presence and properties of expanding ionized shells within luminous \hii\ regions)
within an energetic framework of a set of models, that these ionized shells
are the precursors of the previously observed class of neutral shells which have
larger radii and lower expansion velocities. Although the scenario 
we present here is conventional in the sense that it links the 
HI shells to OB associations, our quantitative 
treatment is useful in showing that observations such as those of 
Hatzidimitriou et al. (2005) are in fact compatible with it. 
Typical measured velocities of HI shell expansion (e.g. Walter \& Brinks 1999), 
are of order 10~\kms, much lower than the velocities detected 
in OB stellar winds (Herrero et al. 1992) and very much lower 
than velocities in supernova remnants (Weiler \& Sramek 1988, 
Chevalier 1977). The expansion velocities we find for the \ha\ shells 
are of order of 100~\kms. In the context of our schematic model 
developed in this article these velocities are characteristic 
of the successive stages of the developing shell as it expands 
and does work against its surroundings. The timescales implied 
for a shell to reach radii of order 1~kpc are at minimum a few 
times 10$^7$yr, which gives time for the stars in a coeval OB 
association to disappear. We also note that an expansion occurring 
in an anisotropic zone of the interstellar medium (ISM) will 
expand preferentially in the low density direction, so that an 
initiating star cluster would not lie close to its centre; for 
the largest shells this effect would be enhanced by Galactic 
differential rotation.

For these reasons we set out to detect the counterparts of HI 
shells at a much earlier stage, while they are more compact 
but expanding at higher velocities. Of course this implies 
that we cannot be looking at the same objects as those detected 
{\it via} the HI shells and holes, but rather at the \hii\ regions which 
surround the OB associations and which mark the energetic impact 
of these associations on the ISM. To do this we used the optical 
technique most akin to 21~cm line mapping of HI: two dimensional 
emission line mapping in \ha\ with a Fabry-Perot interferometer. 
Some of these results have been reported previously (Rela\~no \& 
Beckman, 2005a, Paper I), and showed the presence, within 
luminous \hii\ regions, of dominant expansive components whose 
properties indicated that they could be the precursors of large HI 
shells. In section~2 of the present article we summarize those 
observations, and describe further work of the same basic type 
on \hii\ regions in the nearby galaxy NGC 5194 (M51).
In section~3 we describe a very simple dynamical model in which the 
inputs from stellar winds and from supernovae are coupled to the 
surrounding gas, taking energy and momentum conservation as basic 
constraints. We apply this to our observed HII regions, using the 
measured \ha\ luminosities to estimate the massive stellar content 
used to calculate the energy and momentum inputs to the shells. 
We go on to compare these results with the range of energies and 
momenta in HI shells from the literature. In section~4 we discuss 
general physical considerations relevant to modelling the implied 
evolutionary sequence and give our conclusions.

\section{Evidence for expanding shells in luminous HII regions}

Using galaxies whose populations of \hii\ regions have published photometric
catalogues in \ha\ (Rozas \etal 1996; Rozas \etal 2000; Rela\~no et al. 2005b)
we used complete intensity-velocity maps, also in \ha, made with
the TAURUS-II Fabry-Perot system on the 4.2m WHT (Paper I) to make an exhaustive study of the integrated emission line 
profiles of the \hii\ region population in three spiral galaxies: NGC~6951, NGC~3339 and
NGC~1530. We found that more than one third of the regions in each galaxy
showed high velocity low intensity peaks in the wings of the main emission
lines. The fraction of the line profiles showing these {\it wing features}
increased with the S:N of the observations, so that, for example, NGC~1530, the
galaxy with the highest S:N was the galaxy with the highest fraction of its
profiles exhibiting wing features.

The interpretation we gave (Paper I) to these low intensity high
velocity features is that they are evidence for supersonic expanding shells
within the \hii\ regions, and we could use their observational parameters to
derive the properties of the shells. Fitting the features with Gaussians we
could derive, first, the emission measure corresponding to the feature, which
was used to derive the column density of a given shell and, using reasonable
assumptions, go on to estimate its density (see Eq. (1) of Paper I). 
Secondly, the mean velocity separation between the central Gaussian emission 
component and the lower intensity wing features gave us the velocity of the expanding 
shell.

Here we have taken the dynamical parameters (mass and kinetic energy)
of a sample of \hii\ regions from Paper I. The sample was 
selected to cover the full range of \ha\ luminosities of the \hii\ regions observed
in the three galaxies (see Table~\ref{table_ha_shells}). The estimated shell radii shown in
Table~\ref{table_ha_shells} are rather crude estimates, since at the distances of the galaxies
the linear resolution does not allow a clear measurement. Based on previous
studies of expanding shells in extragalactic luminous \hii\ regions much 
closer to the observer, 30 Doradus (Chu \& Kennicutt 1994) and NGC~604
(Yang et al. 1996), in Paper I we used a canonical shell radius of $\rm {R_{shell}=0.2~R_{reg}}$
(i.e. 20\% of the \hii\ region radius) and a shell thickness of 4.5~pc. From
these estimates we could go on to estimate the shell mass as a fraction 
of the total gas mass in the \hii\ region. Average values for the \hii\ regions in the
three galaxies ranged from 20\% for NGC 1530 to 26\% for NGC 3359, i.e. of 
order 20-25\%. We could then use the estimated mass and the measured velocity to
derive an estimated kinetic energy of expansion for each shell. We obtained 
the equivalent number of O3(V)
stars using the \ha\ luminosity of a given region and the value given by Vacca et 
al. (1996) for the ionizing photon output of an O3(V) star. We can later use the 
equivalent number to estimate the kinetic energy input from the stellar winds, 
using the estimate by Leitherer (1998) of the wind luminosity for an O3(V) star,
integrated over the time the star is on the main sequence. This input energy
is listed in the last column of Table~\ref{table_ha_shells}. It is clear that we could have
added the extra sophistication of looking at an IMF, instead of choosing a
specific stellar spectral type, but the uncertainties implied would not have
made this option a better one in terms of accurate quantification of the
phenomena.

\begin{table*}[]
\caption[]{Shell parameters of a subsample of \hii\ regions in
  NGC~1530, NGC~3359, NGC~6951 and NGC~5194. The shell parameters of
  the \hii\ regions in the first three galaxies are from Paper I, while
  the rest are derived here.
Column 1: Number for the \hii\ region in the catalogue of each galaxy. Column 2:
  Logarithmic \ha\ luminosity. Column 3: Shell 
radius. Column 4: Measured shell expansion velocity. 
   Column 5: Shell electron density. Column 6: Ionized mass of the shell. 
   Column 7: Kinetic energy of the shell. 
Column 8: Combined kinetic energy of OB stellar winds, assuming an
equivalent number of O3(V) type stars.}
\centering
\begin{tabular}{cccccccc}
\hline
Region & log~L$_{\scriptsize\ha}$ &  $\rm R_{shell}(\ha)$ & $\rm v_{shell}(\ha)$ & $\rm n_{shell}(\ha)$ 
& $\rm M_{shell}(\ha)$ & $\rm E_K(\ha)$ & $\rm E_{wind}(O3)$ \\ 
 (number) & (\ergs) &  (pc) &  (\kms) & (\cmtres) & ($\rm 10^{4}\Msun$)  
& ($10^{51}$~erg) &  ($10^{51}$~erg) \\
\hline
NGC~1530 & & & & & & &  \\ 
\hline
8  & 39.45 & 82.00 & 64.72 & 10.90 & 10.2  & 4.3 & 10.4 \\
22 & 39.12 & 60.92 & 49.38 &  9.37 & 4.9   & 1.2 &  4.9 \\
92 & 38.29 & 27.24 & 79.13 &  9.02 & 0.9   & 0.6 &  0.7 \\
\hline
NGC~3359 & & & & & & & \\ 
\hline
6   & 39.15 & 69.62 & 63.97 &  8.36 & 5.7 & 2.3  &  5.3 \\
42  & 38.58 & 48.66 & 40.91 &  8.16 & 2.7 & 0.5  &  1.4 \\
92  & 38.19 & 36.38 & 51.98 &  4.80 & 0.9 & 0.2  &  0.6 \\
\hline
NGC~6951 & & & & & & &  \\ 
\hline
2  & 39.27 & 66.62  & 60.93 & 11.11 & 6.9 & 2.5 & 6.9 \\
18 & 38.67 & 48.66  & 47.62 &  6.50 & 2.2 & 0.5 & 1.7 \\
41 & 38.35 & 37.34  & 50.82 &  8.27 & 1.6 & 0.4 & 0.8 \\
\hline
NGC~5194 & & & & & & &\\ 
\hline
403 &38.94 & 63.93 & 46.75 &  8.96 & 5.1 & 1.1 & 3.2 \\  %(13)%
312 &38.66 & 35.30 & 56.75 & 10.02 & 1.7 & 0.6 & 1.7 \\ %(15)%
416 &38.46 & 52.48 & 50.80 &  4.08 & 1.6 & 0.4 & 1.1 \\ %(18)%
\hline
\end{tabular}
\label{table_ha_shells}
\end{table*}

\subsection{Observations of NGC~5194}

We have followed the method used in Paper I and outlined in 
the paragraph above to study a sample of \hii\ regions in the nearby galaxy NGC~5194 (M51).
The \ha\ Fabry-Perot observations for this galaxy were obtained
  on the 1.6~m telescope of the Observatoire du
Mont M\'egantic (OMM) using the instrument F{\sc {a}}NTO{\sc {m}}M
(Hernandez et al. 2003; Gach et al. 2002). The final data cube
consists of 48 spectral channels with a corresponding total
integration time per channel of 5.16~min. The channel resolution is 0.15\AA\ (corresponding to
6.94\kms) and the effective resolution of the observations is R=18609,
which corresponds to 16~\kms. The central wavelength of the redshifted narrow band blocking filter for \ha\
is 6581\AA, with a half width of 9.9\AA\ corresponding to 450~\kms. 

The Photon Counting Camera had a scale of 1.6\arcsec/pixel and the seeing during the 
observations was 3-4\arcsec. Although this seeing cannot be considered as very good, the
high spectral resolution allows us to study well the high velocity 
features in apertures commensurate with the seeing, which allows us to study the largest
and most luminous \hii\ regions adequately. A detailed description of the
observations of NGC~5194 is given in Daigle et al. (2006a) and the
data reduction is explained in more detail in Daigle et al. (2006b). The
\ha\ luminosities for the \hii\ regions in NGC~5194 were taken from a 
catalogue obtained by Rand (1992), and kindly supplied by the author. The radius of a
region found in the catalog was used to estimate the highest aperture
for the line profile of any given region under study.

\begin{figure*}
   \centering
   \includegraphics[width=8.7cm]{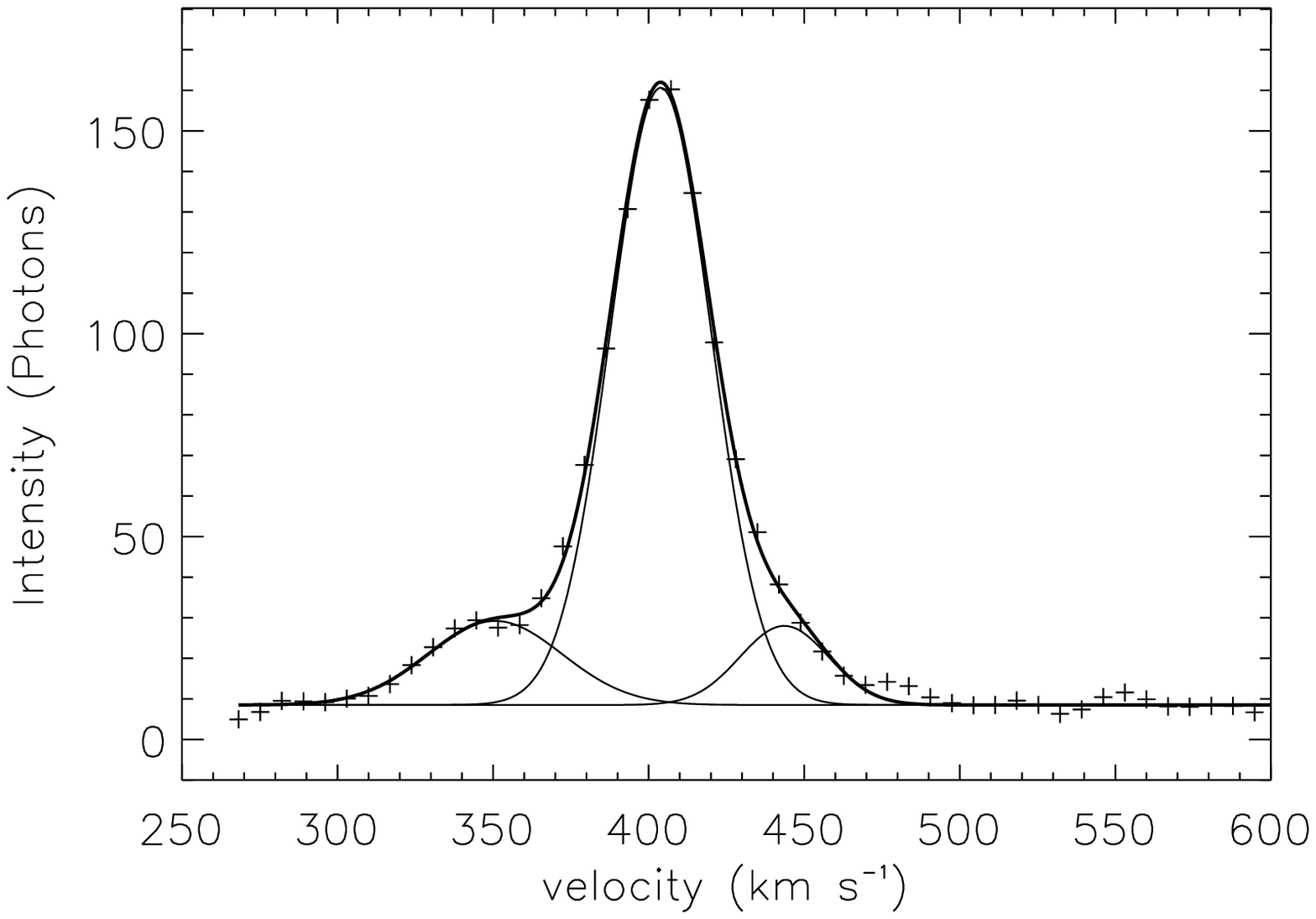}
   \includegraphics[width=8.7cm]{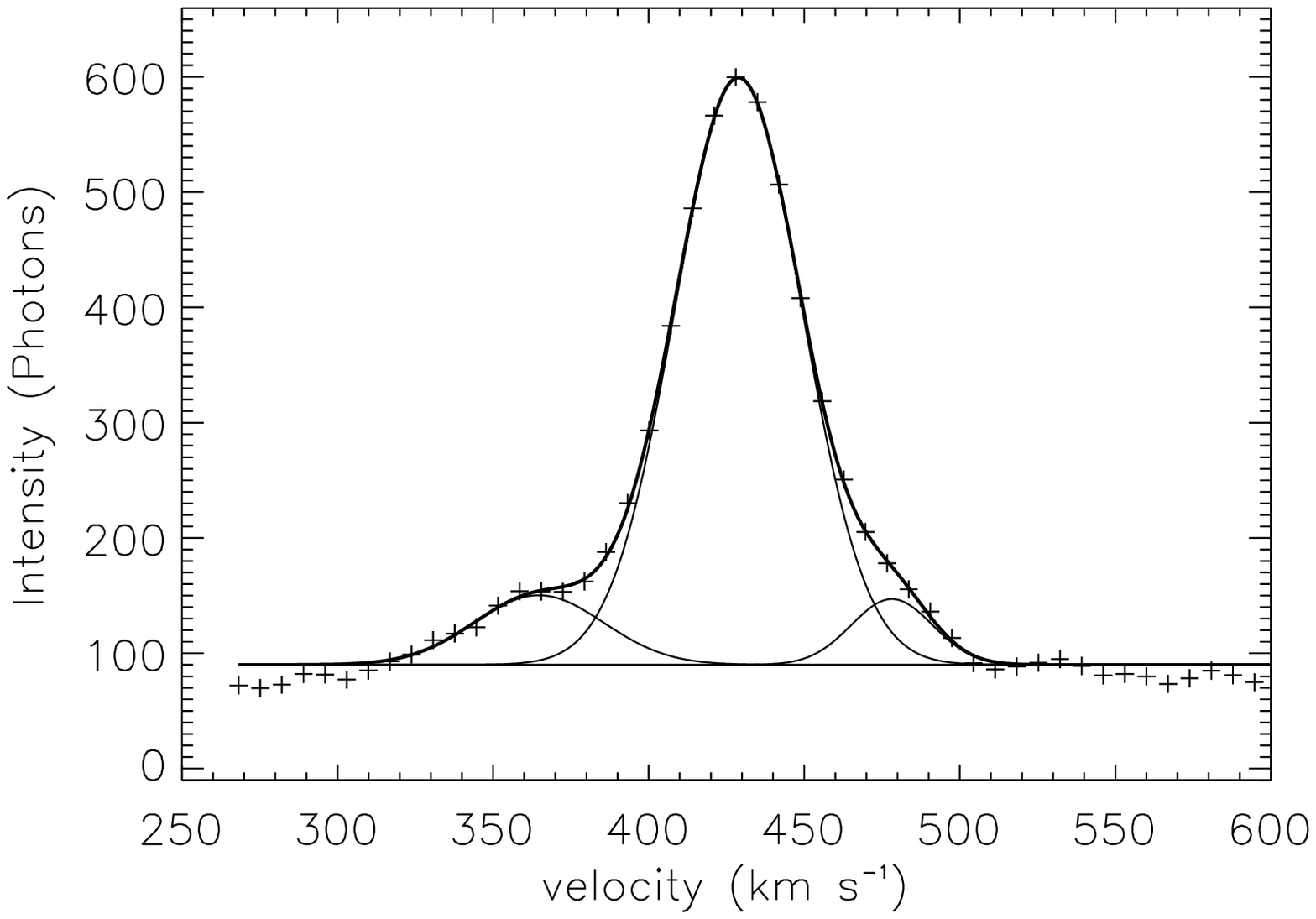}
   \caption{Left: Emission line profile in \ha\ of the \hii\ region 403 of M~51 in Rand's photometric
catalogue. The profile was obtained from the Fabry-Perot data cube of
the \hii\ region. The Gaussian decomposition shown
in the figure was obtained using the task {\sc PROFIT} in the GIPSY
package (van der Hulst et al. 1992). Right: Emission line profile
   of the \hii\ region 312 of M~51 in Rand's catalogue obtained and fitted in
   the same way as in the left figure.}
          \label{espectro}
    \end{figure*}

In Fig.~\ref{espectro} we show two examples of line profiles for \hii\
region 403 and 312 of M51 in Rand's catalogue with the corresponding Gaussian
components fitted to this spectrum. These figures clearly show the
typical red and blue components called here {\it wing features}. For an
extended sample of the line profiles from \hii\ regions in other
galaxies the reader is referred to Paper I. The results of the 
analysis for the emission line profiles of the \hii\ region sample in NGC~5194 are shown in
Table~\ref{table_ha_shells}. For each \hii\ region we
extracted line profiles with apertures of increasing radius, and fitted them
with the corresponding Gaussian components. The profile whose wing features
yielded the best S:N was taken as the representative spectrum of the
\hii\ region. The observational parameters of the wing features for
each \hii\ region do not vary significantly when increasing the aperture, which
allows us to take the best S:N line profile as the representative one
for each \hii\ region. The corresponding observational parameters,
the Emission Measure and the
velocities of the wing feature Gaussians with respect to the central 
emission peak were used to obtain the dynamical parameters of the shell for each 
region using the same criteria as those in Paper I. These parameters are 
listed in Table~\ref{table_ha_shells}. The relative errors for the
observational parameters are 0.2-0.3 for the Emission Measures and 0.05-0.08 for the
velocity separation between the central and the shifted components. It
is interesting to note that the absolute errors in the fitting
procedure for the velocity of the shifted
components are 2-3~\kms, much lower than the spectral resolution of
the observations. This gives an idea of the validity of the Gaussian
decomposition procedure.

\section{Energetics of the expanding shells}
\subsection{Starburst99 Models}
\label{st99models}
Our aim here is to extrapolate from the properties of the shells we have
observed within the \hii\ regions to later times, taking the shells as
representative, to show how their parameters would evolve with time
and then to compare the extrapolated shells with HI shells observed in the Galaxy and
nearby galaxies. In order to infer the energy inputs from the stars in the
form of both stellar winds and supernovae, we have used from the 
literature the Starburst99 models (Leitherer et al. 1999). We take the case of an
instantaneous starburst with a fixed stellar mass and integrate using two
different Initial Mass Function (IMF) exponents, 1.3 and 2.3 for mass 
intervals (0.1-0.5)$\rm\Msun$ and (0.5-100)$\rm\Msun$, respectively. We take the
conventional assumption that 8$\rm\Msun$ is the lower limit of a Zero
Age Main Sequence star which 
will eventually explode as a supernova, and use the Padova evolutionary tracks at
solar metallicity. Starburst99 then gives, from an initial time of $10^{4}$yr
the (logarithmic) energy (in erg~s\me) input due to both stellar winds and
supernovae as a function of time.

The principal input parameter in our Starburst99 models is the stellar
mass associated with the burst, which was inferred from the \ha\ 
luminosity of the \hii\ region, as described in section 7.2.1 of
Rela\~no et al. (2005b). We use the Salpeter IMF to estimate the {\it ionizing mass} of
the star cluster and compare it to the mass estimate using an 
equivalent number of O5(V) stars. 
The normalizing factor obtained is then used when integrating the IMF within the 
appropriate mass limits
(0.1$\rm\Msun$-100$\rm\Msun$) and the total stellar mass in the cluster is derived.
The output of Starburst99 gives the energy produced by the stellar winds and
supernovae within the star cluster as a function of the age of the
cluster. We have taken the total energy produced by the stellar winds and also by
the supernovae integrated over a time scale of 1.0$\times 10^{7}$yr, and have studied
the evolution of a shell over this time scale. Estimates of the ages of \hii\ regions 
based on predicted ratios of emission line strength to continuum level 
(Copetti et al. 1985; Bresolin \& Kennicutt 1997) yield ages for \hii\ regions in the 
range from 2Myr to 8Myr, and show that the dynamical timescale adopted here is reasonable.

\subsection{Two basic driving mechanisms: Stellar Winds and Supernova explosions}

There are two basic mechanisms which can drive the formation of a
shell expanding into the ISM from an OB association: the stellar winds from
young stars and the explosions of supernovae. The first mechanism was
modelled when accounting for the shells observed in the \hii\ region 
populations of the three spiral galaxies described in Paper I. We
compared there the wind 
energy input from the equivalent numbers of O3(V) stars, derived using the \ha\
luminosity of a region, with the shell kinetic energy inferred from the
kinematic observations. The conclusion was reached that shells with 
radii of up to $\rm\sim 0.2~R_{reg}$ (where $\rm R_{reg}$ is the radius of
the \hii\ region) could be quantitatively explained by the interaction of the stellar winds with the
ISM, while some of the bigger shells located within the most luminous
\hii\ regions show kinetic energies which are too high to be explained as the
result solely of wind interaction.

We have here gone on to improve the method used in Paper I, and now
use the input energy from the stellar winds (here derived from the models
implemented in Starburst99), but also add in the energy released in
the subsequent supernovae. We are thus modelling the situation in which both
mechanisms operate on the ISM surrounding a star cluster. The winds
provide a continuous energy source which lasts for the stellar
lifetime, i.e. $\sim$10$^{6}$yr, by this time an expanding shell 
has formed within the \hii\ region. This is a single shell even 
if the stars "switch on" at dispersed times, because if the first star 
causes a shell to form, the wind from a second star will
expand more quickly until it reaches the shell, and then couple its
impulse to the shell, and this will occur for each subsequent star.
Then, the supernovae occur inside the partially attenuated bubble produced by
the winds. The situation has been described in some detail by
Bruhweiler et al. (1980) and by Dyson (1981); the shock produced by
the SN explosion sweeps up the material inside the cavity formed by
the stellar winds, and impacts directly on the shell; then the
strength of the transmitted shock is reduced by a factor which depends
on the contrast in density between the cavity and the newly formed shell. The radiative cooling time
within the shell can be very short, and then the momentum conservation is applied
as the best approximation to describe the dynamics of the new formed
shell.

The initial velocity is taken directly from the \ha\ observations, so that 
$\rm v_0$ is the velocity of the shell as observed within an \hii\ 
region, in \ha\ (column 4 of Table~1). The radius at which the SNe 
explode ($\rm R_{0}$), is taken to be the radius of a wind blown 
shell. This is estimated applying the model of Dyson (1980) for 
the adiabatic expansion of a stellar wind bubble. The inputs 
required in this model are the 
energy flux in the stellar winds, the initial gas density, and 
the time during which the stellar winds act. Taking the stellar wind 
energy from Starburst99 models we apply the equations of the adiabatic model 
to a range of values of n and $\rm t_{0}$. The optimum combination 
of these variables for producing the measured parameters of the 
wind blown portion of the shell expansion is n=0.5\cmtres\ and $\rm t_{0}=10^6$yr. 
Using these values the shell radius is found to be $\sim 0.3\rm R_{reg}$ and 
the velocities just slightly lower than the observed \ha\ velocities. 
In the end we chose to use $\rm R_{0}=0.5R_{reg}$ for the initial 
radius of the shell in the post SN phase of our modelling. This is the 
shell radius at which the SNe explode and although it is a little
higher than the radius derived in the adiabatic model, it allows us to 
simplify our modelling by separating the adiabatic expansion phase 
due to the winds from the momentum conserving phase after the 
supernovae have exploded. We used an ambient interstellar density of n=0.1\cmtres, 
slightly lower than the value estimated from the adiabatic model. 
This values is chosen as a practical compromise which takes 
partially into account that the density decreases with distance 
from the plane of the galaxy, a circumstance of relevance especially 
for the larger HI shells. We confirmed that using a density of 
0.5\cmtres\ changes the magnitude of the results by a relative modest fraction 
(slightly ($\sim$20\%) higher radii and higher 
velocities ($\sim$13\%)).

\begin{figure*}
   \centering
   \includegraphics[width=8.7cm]{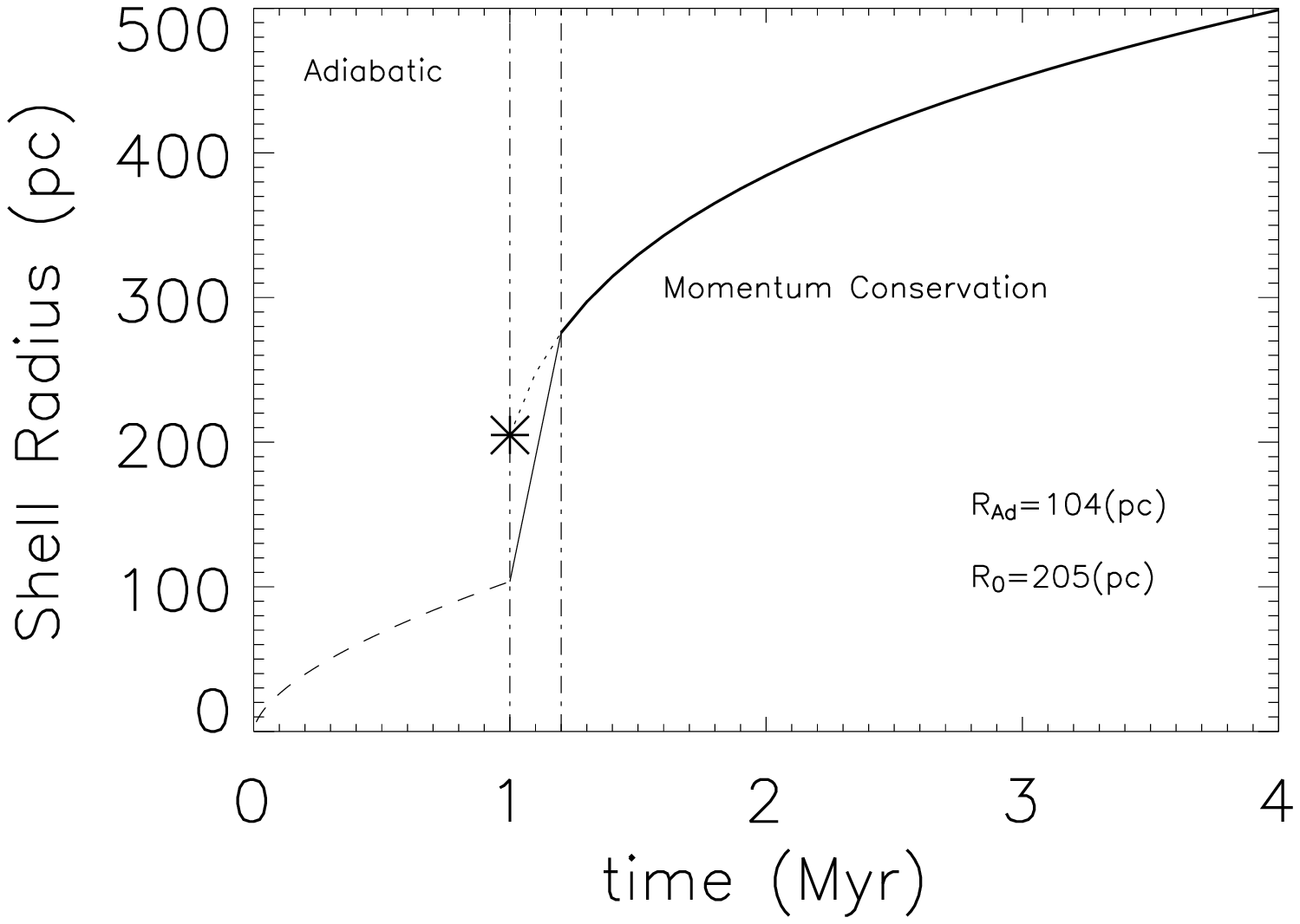}
   \includegraphics[width=8.7cm]{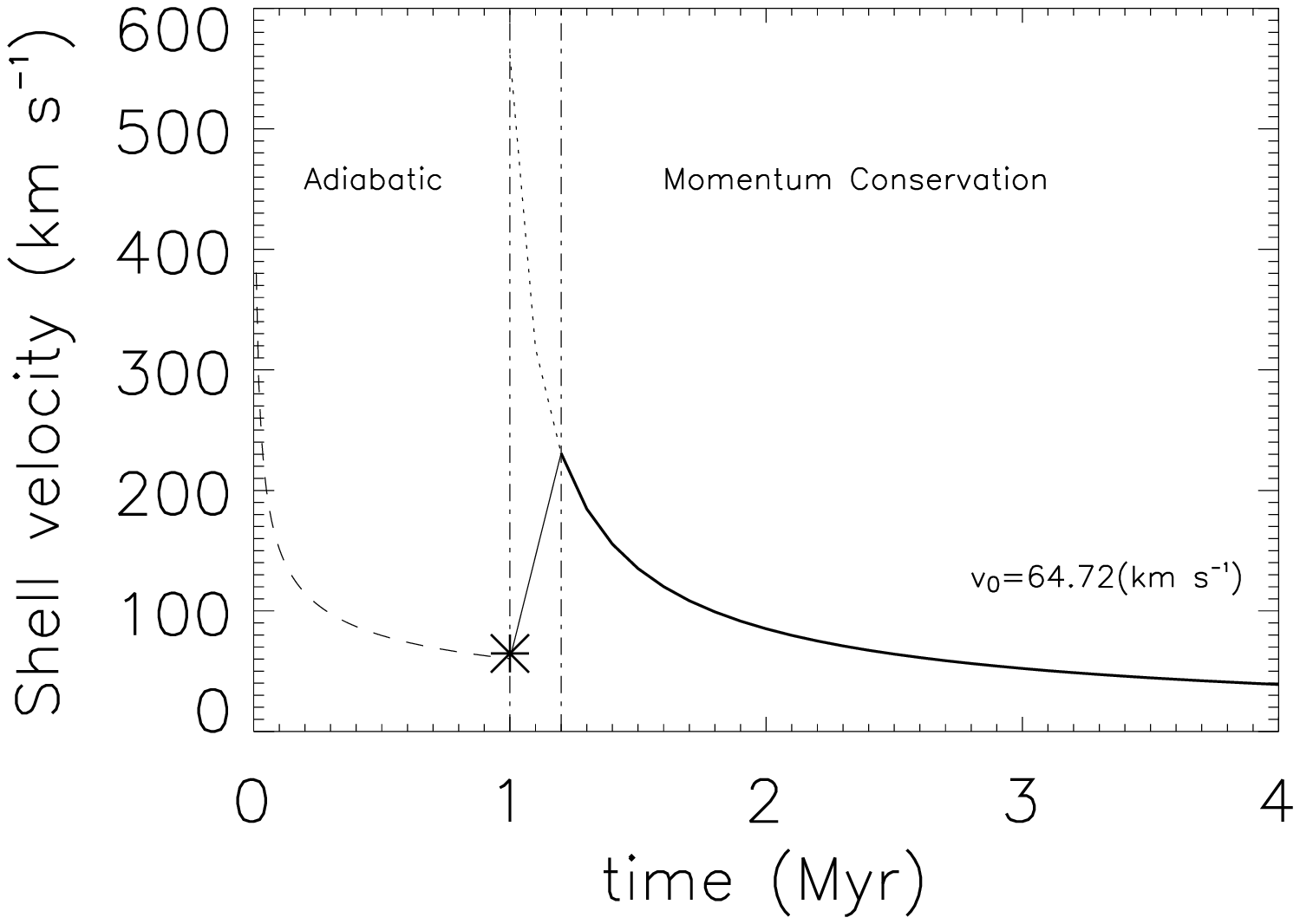}
   \caption{Left. Evolution of the shell radius of the adiabatic and momentum conserving phases
for the parameters shown by \hii\ region~8 of NGC~1530. R$_{\rm Ad}$ is the 
initial radius found from the adiabatic model, which predicts in general shell radii of 
0.3R$_{\rm reg}$, and velocities comparable with those observed for the ionized shells 
within the \hii\ regions. R$_{\rm 0}$ is the input radius for the momentum
conserving phase. We have not explicitly modelled the transition between the phases, and this
is represented schematically here by the line joining the two curves in the time 
interval between 1 and 1.2~Myr.
Right. Evolution of the shell velocity for the same example \hii\ region. 
The shell velocity predicted in the adiabatic phase is close to the observed velocity of the \ha\ 
shell (see column~4 of Table~\ref{table_ha_shells}). In our simplified model 
of the momentum conserving phase, in 
which the SN input is instantaneous, the initial velocity  would be boosted to the 
unrealistically high value indicated in the figure. A schematic representation of 
the transition is given by the line joining the two curves in the time interval 
between 1 and 1.2~Myr.  Although the assumption of instantaneous
energy injection is too simple, the overall energetics of the model, which predicts a 
final HI radius of 653.5~pc and a corresponding HI expansion velocity of 17~\kms\ 
(see Table~\ref{table_energy_HI}) are physically reasonable}.
          \label{plot_model}
    \end{figure*}

Using the model assumptions described above we start with the shell whose
properties we have derived from our \ha\ observations and run the models
adding the supernova inputs as explained. We assume that the
stellar winds have acted over a time interval of $\rm t_o=10^{6}$yr 
and that at this time the supernovae explode with no time intervals between 
them. The supernovae then sweep up the material within the bubble and push
outwards the shell created by the stellar winds. The application of momentum
conservation gives the radius and the shell velocity as a function of
time according to the expressions (Dyson 1980; Bruhweiler et al. 1980):

\begin{equation}
{\rm R_{shell}(t)=[R_{0}^4+\frac{3M_{0}v_0}{nm_{H}\pi}
(t-t_{0})]^{(1/4)}}
\label{eqrad}
\end{equation}

\begin{equation}
{\rm
  v_{shell}(t)= R_{shell}^{-3}\frac{3M_{0}v_0}{4nm_{H}\pi}}
\label{eqvel}
\end{equation}
where $\rm R_{0}$ and $\rm v_0$ are the radius and velocity of the
expanding shell at the switch-on time of the supernovae, n is the
ambient interstellar density and $\rm M_{0}$ is the mass swept up 
by the effect of the supernova explosions. In Fig.~\ref{plot_model} 
we show the time evolution of the shell radius and shell velocity
for the \hii\ region~8 in NGC~1530 (first raw of Table~\ref{table_ha_shells}).
The figure, that can be taken as a characteristic example, shows 
qualitatively how the shell forms in the wind blown phase of the 
expansion. It then receives a boost from the supernova input which
initially accelerates it, but the velocity is subsequently reduced as 
the shell picks up mass from the surrounding ISM and expands reaching a 
radius within the range measured for HI supershells.

The mass $\rm M_{0}$ in Eq.~\ref{eqvel} is the total mass contained within the 
bubble at the initial radius $\rm R_{0}$, which is assumed to be
completely swept up as a consequence of the SN explosions. To obtain 
the mass within the bubble 
we have used the method to derive the mass of gas in an \hii\ region
described in Rela\~no et al. (2005b), where we included a factor to
take into account the neutral gas
not observed directly in \ha\ (see Giammanco et al. 2004).
We then apply Eqs.~1 and~2 and compute a "final" radius for the shell, $\rm R_{shell}$, and a
"final" velocity, $\rm v_{shell}$, at a time of $1.0\times 10^{7}$yr,
which agrees with the dynamical age range of the suppershells
observed in HI (McClure-Griffiths et al. 2002; Walter \& Brinks
1999). An integration of the models at a time of $0.5\times 10^{7}$yr
will produce velocity values between 15-30\kms, while at longer times 
($5.0\times 10^{7}$yr) the velocities will be $\sim$3\kms, much lower than
those shown here. These last values are not expected to have a clearly
defined physical meaning, since the shell loses its identity at v$\sim$5\kms\ 
(Dyson 1981), and for the corresponding radii, of order 500~pc of greater, a shell will tend to 
be disrupted by global galactic effects such as differential rotation.
We present the results of the models in Table~\ref{table_energy_HI}, 
where we also show the total
combined input energies of stellar winds and supernovae from
Starburst99, and the kinetic energy of the bubble obtained using the
standard expression $\rm E_k=\frac{1}{2}m_{HI}v_{HI}^{2}$. 
An interesting supplementary 
exercise was to run models in which 
a second burst of star formation boosted the injection of energy into 
the expanding shells. According to the time at which the 
model output is examined this produces HI shells either 
with a range of radii similar to those observed but with 
a velocity range substantially higher, or with a range of 
velocities similar to those observed but with a range of 
radii substantially higher. This suggests strongly that 
the stars in OB associations are coeval. In the
next section we compare our results with a sample of HI shells
observed in two different galaxies.

\begin{table*}[]
\caption[]{Energetics of the HI predicted shells for the sample of \hii\
  regions in NGC~1530, NGC~3359, NGC~6951 and NGC~5194. The stellar winds 
act over $10^{6}$yr, the time at which the SNe explode, the final integration time
of the models is $10^{7}$yr. Column 1: \hii\ region number. Column 2: Shell 
radius. Column 3: Shell expansion velocity. Column 4: Mass of the shell.
   Column 5: Energy input from the supernovae and stellar winds. 
Column 6: Kinetic energy of the shell. Column7: Efficiency of the
  SN to convert their energies into kinetic energy in the ISM.} 
\centering
\begin{tabular}{ccccccc}
\hline
Region &  $\rm R_{shell}$ & $\rm v_{shell}$ & $\rm M_{shell}$ & $\rm
E_{(SW+SN)}$ & $\rm E_{K}(shell)$ & Efficiency \\ 
 (number) &  (pc) & (\kms) & ($\rm 10^{5}\Msun$) & ($10^{52}$~erg) &
($10^{51}$~erg) & $(\%)$ \\
\hline 
 1530-8 &  653.5 & 17.3 & 28.9 &  36.1 & 8.6 & 3.7 \\
1530-22 &  505.8 & 13.4 & 13.4 &  16.8 & 2.4 & 2.2 \\
1530-92 &  351.6 &  9.4 &  4.5 &   2.7 & 0.4 & 2.3 \\
\hline
 3359-6 &  550.2 & 14.6 &  17.2 & 18.1 & 3.7 & 3.2 \\ 
3359-42 &  353.5 &  9.3 &   4.6 &  4.7 & 0.4 & 1.3 \\
3359-92 &  299.7 &  8.0 &   2.8 &  2.0 & 0.2 & 1.4 \\
\hline
6951-2  &  579.4 & 15.4 &  20.1 & 24.1 & 4.8 & 3.1 \\
6951-18 &  386.8 & 10.3 &   6.0 &  6.0 & 0.6 & 1.6 \\
6951-41 &  326.9 &  8.7 &   3.6 &  2.7 & 0.3 & 1.6 \\
\hline
5194-403 & 449.7 & 11.9 & 9.4 &  10.5 & 1.3 & 2.0 \\%(13)%  
5194-312 & 400.9 & 10.7 & 6.7 &   5.5 & 0.8 & 2.1 \\%(15)%  
5194-416 & 348.8 &  9.2 & 4.4 &   3.7 & 0.4 & 1.5 \\%(18)%  
\hline
\end{tabular}
\label{table_energy_HI}
\end{table*}

\subsection{Comparison with HI observations}

We have selected a representative sample of HI shells with different sizes 
observed by Walter \& Brinks (1999) in the dwarf galaxy IC 2574 and by 
McClure-Griffiths et al. (2002) in the Milky Way. We have limited our
sample to those shells classified by Walter \& Brinks as "classic holes"
and we have avoided "merged shells" and chimneys in the McClure-Griffiths 
et al. data set. Our results in Table~\ref{table_energy_HI} show good broad agreement with the 
parameters of the observed HI shells reported in the two articles cited. The modelled
range of shell velocities, $\rm v_{shell}=(8-17)$\kms, falls nicely within the
range of the observed sample, $\rm v_{shell}=(4-21)$\kms. The range of shell 
masses produced in the models, $\rm M_{shell}=(3-29)\times
10^{5}\Msun$ is within the range
of masses observed by McClure-Griffiths et al.(2002), $\rm (1-44)\times 10^{5}\Msun$ but is
somewhat higher than the range found by Walter \& Brinks (1999), $\rm (0.02-18)\times 10^{5}\Msun$.
The first study yields values for
the HI number density 0.5-1.7\cmtres, while the second study takes the scale height of the
galaxy disk into account (the galaxy is in any case much smaller) and finds
values in the range 0.05-0.17\cmtres. 
The smaller values for the range of the basic parameters in Walter \& Brinks
(1999) can be ascribed to the fact that they were observing a dwarf galaxy
where the range of ISM densities is lower, and this will necessarily
be reflected in the ranges of shell mass and kinetic energy. Our \hii\ region
observations have all been taken from regions in essentially Milky Way mass
galaxies. The better agreement between the range of parameters derived by
applying dynamical modelling to shells observed in these regions, using the
appropriate values of mass, velocity and ISM density, and the HI shells
observed in the Milky Way by McClure-Griffiths et al. is not
fortuitous. 

%\section{Some physical considerations}
\section{Final Remarks}

It is not surprising to find expanding shells around OB associations,
but a variety of scenarios have been proposed in which these can
arise. OB stars have a major dynamical impact on their surroundings,
which can take four forms: the direct influence of ionizing radiation 
on the surrounding gas, the influence of non--ionizing radiation on
the surrounding gas, mediated by its dust content, the effect of the
outflow of gas from the stars themselves in the form of stellar winds,
and the effects of the violent outflow of gas in supernova
explosions. Our underlying concepts here are based on the seminal 
article by Dyson (1981), who considered the dynamical
effects of stellar winds on \hii\ regions in the necessary presence of the
ionizing radiation which forms them. 

In treating the development of the wind-blown shells we must then take
into account that on a well determined timescale there will be an
additional dynamical input from the supernovae in the OB
association. The timescale is of key importance, because during the
main sequence lifetime of an O star, of order $10^6$yr, a wind--blown
shell expanding at a characteristic velocity of $\sim$50\kms\ will reach a radius of $\sim$50~pc. This
implies that the supernovae, which explode in the star cluster will at
first expand very rapidly into a low density environment cleared by the 
winds, but will then impinge on the denser region marked by the shell and the
surrounding ISM. Under these circumstances the adiabatic phase
(Sedov 1959; Chevalier 1977)
of the SN remnant expansion will immediately give way to the snowplough phase 
(Chevalier 1974) and the dynamical coupling of the SN remnant to the expanding shell will be
determined by momentum conservation. We find that a typical 
fraction of the expansion energy of the SN remnant transferred to the shell will be 
(1-4)\%\ (see last column in Table~\ref{table_energy_HI}), with the
rest either taken up as turbulence or radiated away. Further 
observations and more complete analysis would be necessary in order to set more rigorous constraints on
the efficiencies under a variety of conditions. For example, if the largest 
bubbles blow out when expanding into lower gas densities, the efficiency of the 
process will be predictably lower than those found here.

To summarize, the detection and measurement of 
the physical parameters of dominant expanding shells 
within a major fraction of highly luminous extragalactic 
“giant” \hii\ regions presented in Paper~I and fully 
confirmed by the observational results in the present article, 
has given us a starting point from which to present a scenario 
relating the presence of these shells to the presence and the 
properties of the much larger HI shells detected in the Milky Way 
and in nearby galaxies. The result of applying a straightforward 
model embodying this scenario is that the HI shells in general 
most probably had \ha\ shells as their precursors. The time scales 
within the model are such that one should not expect to observe both 
the precursor shell and the HI shell for the same system, so that 
our results are necessarily statistical in nature. However we 
can conclude that the observations of Hatzidimitrou et al. (2005), who 
observed stars in OB associations at the centres of only a minority 
of their HI shells, are fully compatible with our scenario. A suggestive, 
though perhaps not conclusive secondary inference, is that the stars 
in an OB association probably form in a single group not 
dispersed in formation epoch.

\begin{acknowledgements}
This work has been supported by the Spanish Ministry of Education and Science
within the PNAYA (Spanish National Programme for Astronomy and Astrophysics)
via projects AYA2004-08251-CO2-00, and ESP2003-00915, and by project
P3/86 of the Instituto de Astrof\'\i sica de Canarias. We thank
Richard Rand for kindly providing us with the \hii\ region catalogue of
NGC~5194. We are grateful to John Dyson for valuable discussions on 
theoretical aspects of this work. We thank the second anonymous 
referee for helpful comments and suggestions which
enabled us to improve the article.
\end{acknowledgements}

\end{document}